\documentclass[%
 reprint,
 showpacs,preprintnumbers,
 amsmath,amssymb,
 aps,
 prb,
]{revtex4-1}
\usepackage{dcolumn}
\usepackage[dvipdfm]{graphicx}
\usepackage{subfigure,color}
\usepackage{mathrsfs}

\makeatletter
\def\btt#1{\texttt{\@backslashchar#1}}
\DeclareRobustCommand\bblash{\btt{\@backslashchar}} \makeatother

\begin{document}

\title{Gapped energy spectra around the Dirac node 
at the surface of a 3D topological insulator 
in the presence of the time-reversal symmetry}

\author{Tetsuro Habe$^1$ and Yasuhiro Asano$^{1,2}$}
\affiliation{$^1$Department of Applied Physics,
Hokkaido University, Sapporo 060-8628, Japan}
\affiliation{$^2$Center for Topological Science \& Technology,
Hokkaido University, Sapporo 060-8628, Japan}

\date{\today}

\begin{abstract}
We discuss the excitation spectra around the Dirac node on a surface of a three-dimensional topological insulator.
By using the diagrammatic expansion, we show that the coupling of an electron with the gauge field in the presence of impurity scatterings opens a gap around the Dirac node. 
The results are consistent with a recent experimental finding 
by T. Sato, et al. [Nature Phys. \textbf{7}, 840 (2011)].
We also discuss the consistency between the present results and the bulk-boundary correspondence 
of the topological insulator in the presence of time reversal symmetry.
The conclusion can be applied any two dimensional massless fermions.
\end{abstract}

\pacs{73.20.At, 73.20.Hb}

\maketitle

\section{introduction}
Properties of metallic states at the surface of topologically non-trivial 
materials are a hot issue in the condensed matter physics.
A three-dimensional(3D) topological insulator (TI) hosts  
a metallic state with linear dispersion at its surface as a result of 
the topologically non-trivial nature of the wave function in bulk insulating 
region~\cite{Fu2007,Fu2007-2,Moore2007,Chen2009}. 
According to the shape of the dispersion,
such metallic state is called Dirac cone.
A point in the momentum space at which the upper cone and the lower one touch with each other
is called Dirac node. The bulk-boundary correspondence suggests the presence of a single 
Dirac node at the surface, which reflects a topological number $Z_2=1$ in the bulk insulating region. 
The Dirac node is considered to be fragile against perturbations which break the time reversal symmetry (TRS)
because $Z_2$ is well defined only in the presence of TRS. In fact, attaching a ferromagnetic insulator
onto a TI~\cite{Tanaka2009,Habe2012} or introducing magnetic impurities into a TI~\cite{Liu2009,Chen2010} 
remove the Dirac node from the surface state. On the other hand,
the Dirac node is believed to be robust against perturbations which preserve TRS.
The Dirac node remains even in the presence of normal impurity scatterings, which is 
an important feature of the surface state from theoretical, experimental and applicational view points.
 
A recent experiment~\cite{Sato2011}, however, has reported the gapped excitation spectra 
in the single Dirac cone at the surface of $\mathrm{TlBi_2(Se_{x}S_{1-x})_3}$.
Starting with a topological insulator $\mathrm{TlBi_2Se_{3}}$, the gradual substitution of Selenium by Sulfur  
enables to make a series of insulators ended with
a topologically trivial insulator $\mathrm{TlBi_2S_3}$.
The insulators should be topological and should have a gapless single Dirac cone at their surface 
in the doping range of $0.5< x \leq 1.0$. 
The experimental results of the angle resolved photo emission, however, clearly shows the gapped 
excitation spectra in the corresponding doping range. 
At present, we do not have any reasonable argument which explains the  
discrepancy between the theoretical prediction and the experimental results.

Motivated by the experiment~\cite{Sato2011}, we theoretically try to make clear
a mechanism which generates the gap around the Dirac node in the presence of the TRS.
We consider the massless fermion in two-dimension which couples with the gauge field 
in the presence of normal impurity scatterings.
The self-energy due to the impurity scatterings and the coupling with the gauge field
is calculated within the lowest order of perturbation expansion.
The results show that the real part of the self-energy for the upper Dirac cone has the opposite 
sign to that for the lower cone. As a result, two-dimensional fermion has gapped energy spectra 
around the Dirac node.
We conclude that the interplay between the impurity scatterings and 
the dynamical electron-electron interactions via the gauge field generates the gap.
The results suggest the absence of gapless surface state even in the presence of TRS.
We will show that our conclusion is not inconsistent with
the bulk-boundary correspondence of a TI.
We also conclude that the translational symmetry is necessary to preserve the gapless surface 
state in real TIs.

This paper is organized as follows. In Sec.~II, we explain our theoretical model.  
The self-energy due to the coupling with the gauge field in the presence of 
the impurity scatterings is calculated in Sec.~III. We discuss a relation between 
theoretical results 
and an experimental one in Sec.~IV. The conclusion is given in Sec.~V.

\section{Theoretical model}
Let us consider 
the Lagrangian density which describes the two dimensional massless Dirac fermions coupling with the gauge field,
\begin{align}
{\mathscr{L}}=\sum_{j=0,1,2}\;\psi^{\dagger}(\tilde{p}_j-eA_j){\sigma}^j\psi-\frac{1}{4}F_{\mu\nu}F^{\mu\nu},\label{Hamiltonian}
\end{align}
where $\sigma^0$ is the $2\times2$ unit matrix, ${\sigma}^{j}$ for $j=1-3$ are the Pauli matrices,
$\psi$ and $\boldsymbol{A}$ are the field operators of a surface massless fermion and a $U(1)$ gauge boson.
In Eq.~(\ref{Hamiltonian}), $p_0=\tilde{p}_0$ represents energy measured from the Dirac node and $p_j=\tilde{p}_j/v$ for $j=1-2$ represents the momenta multiplied by 
the speed of light $c$ where $v=v_F/c$ is a small constant describing dimensionless velocity with 
 the Fermi velocity $v_F$. 
Throughout this paper, we use the units of $\hbar=c=1$.
The field tensor $F^{\mu\nu}$ is defined by
\begin{align}
F^{\mu\nu}={\partial}^{\mu}A^{\nu}-{\partial}^{\nu}A^{\mu},
\end{align}
with ${\partial}^{\mu}={\partial}/{\partial x_{\mu}}$.
We consider the Dirac propagator defined by
\begin{align}
G_0(p)=\frac{p_0{\sigma}^0+v\boldsymbol{p}\cdot{\boldsymbol{\sigma}}}{{p_0}^2-{v}^2{\boldsymbol{p}}^2+i\varepsilon}, \label{g0}
\end{align}
where we take a short-hand notation $p=(p_0,\boldsymbol{p})$.
Eq.~(\ref{g0}) represents the particle propagator for positive energy $p_0>0$ and the 
hole propagator for negative energy $p_0<0$.
We employ the Feynman gauge in which the propagator converges at the spacial infinity 
$x_{j}\rightarrow\infty$ with $j=1-2$.
The propagator of the gauge field at energy $q_0$ and momentum $\boldsymbol{q}$ is given by
\begin{align}
D_{\mu\nu}(q)=&\frac{-g_{\mu\nu}}{{q_0}^2-\boldsymbol{q}^2+i\varepsilon}, \\
g_{\mu\nu}=&\mathrm{diag}\{1,-1,-1\},
\end{align}
where $\mu$ and $\nu$ are 0, 1, and 2.
As shown in Eq.~(\ref{Hamiltonian}), the vertex of the coupling between the electron and the 
gauge field is described by $-e\sigma^{\mu}$ in Heaviside units. 
We assume that the impurity potential is spin-independent and is represented by the delta-function,
\begin{align}
U_{i}(\boldsymbol{r})=\sum_{j}u{\sigma}^0\delta({\boldsymbol{r}}-\boldsymbol{r}_j),
\end{align}
where $\boldsymbol{r}_j$ is the position of an impurity and $u$ is the strength of a single impurity potential.
We assume a constant number density of impurities $n_i$.

\section{self-energy}

\begin{figure}[tbp]
 \begin{center}
\includegraphics*[height=45mm]{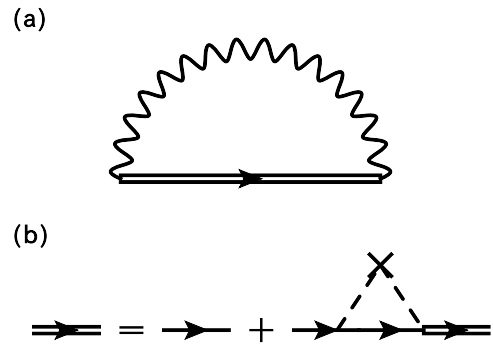}
 \end{center}
\caption{The Feynman diagrams under consideration. 
Figure (a) represents the fermion self-energy coupled to the gauge boson. 
The wavy line denotes the propagator of the gauge boson.
Figure (b) is Dirac fermion propagator in the presence of impurity 
scatterings within the Born approximation.
} \label{fig:SE}
\end{figure}

At first, we estimate the self-energy by the impurity scatterings alone.
The impurity potential mixes the states with different momenta. 
It, however, does not mix the states with different energies.
Therefore impurity scatterings itself cannot open the gap at the Dirac node.
The self-energy within the Born approximation is represented by 
\begin{align}
\Sigma^{(i)}=-i\gamma p_0\sigma^0\label{life-time},
\end{align}
where $\gamma=\pi u^2n_i/{v}^2>0$ is the dimensionless expansion parameter 
smaller than unity (See also Appendix \ref{AP1}).
The result has the general form of the massless Dirac fermion~\cite{Suzuura2002}.

Secondly, we calculate the self-energy of a Dirac fermion coupled with the gauge field which is
given by,
\begin{align}
\Sigma^{(g)}(p)&=
\sum_{q}\sum_{\mu\nu} 
(-e\sigma^{\mu})
G_0(p-q)
(-e\sigma^{\nu})D_{\mu\nu}(q).\label{eq.S1}
\end{align}
The details of calculation are shown in Appendix~\ref{AP2}, where
we estimate the self-energy in Fig.~\ref{fig:SE}(a).
Namely we calculate $\Sigma^{(g)}(p)$ in Eq.~(\ref{eq.S1}) with using the full Green function
$G(p-q)$ in Eq.~(\ref{gimp})
instead of the bare $G_0(p-q)$ in Eq.~(\ref{g0}).
In the limit of small momenta, the real part of the self-energy has an asymptotic form as 
\begin{align}
\mathrm{Re}\left[\Sigma^{(g)}(p)\right]
=\pi e^2\left(\frac{p_0{\sigma}^0}{\sqrt{{p_0}^2}}-\frac{\boldsymbol{p}\cdot{\boldsymbol{\sigma}}}{\sqrt{\boldsymbol{p}^2}}\right). \label{sigma_g}
\end{align}
We note that the result does not contain $\gamma$. Therefore the impurity scatterings add only 
negligible corrections to Eq.~(\ref{sigma_g}) as shown in Appendix~\ref{AP2}. 
From pole of the scalar denominator $[G_0(p)^{-1}-\Sigma(p)]$, this self energy cannot open gap for massless Dirac fermion,
\begin{align}
p_0=&\pm (v|\boldsymbol{p}|+\pi e^2)-\pi e^2\mathrm{sgn}(p_0)\nonumber\\
=&\pm v|\boldsymbol{p}|.\label{unchange}
\end{align}
The two terms stemming from the self-energy 
cancel each other out, which reflects the covariance of the gauge field.
As a consequence, the coupling with the gauge field does not change the 
energy spectra of massless Dirac fermion.
Neither the coupling with gauge field alone nor the impurity scatterings alone 
open the gap at the Dirac node. 
The former changes the energy of a fermion but preserves the translational symmetry.
On the other hand, the latter breaks the translational symmetry and changes the momenta of a fermion.
But it preserves the energy of fermion. 
To have a gap at the Dirac node, we need scattering processes which change 
the momenta and the energy of a fermion at the same time.

\begin{figure}[tbp]
 \begin{center}
  \includegraphics*[height=40mm]{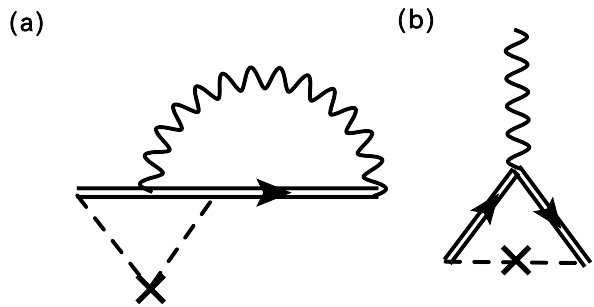}
 \end{center}
\caption{The Feynman diagrams we consider in the text. 
Fig.(a) is the fermion self-energy consisting of the gauge boson and the impurity scatterings.
Fig.(b) represents the vertex function including the impurity scatterings.
}\label{fig:Self}
\end{figure}

Finally, we consider a self-energy which represents the interplay between
the coupling with the gauge field and the impurity scatterings.
The scattering processes are diagrammatically described in Fig.~\ref{fig:Self}(a).
Such self-energy is given by
\begin{align}
\Sigma(p)
&=\sum_{p_0',q,\mu,\nu}{\Gamma}^\mu(p_0',q) G(p-q) (-e{\sigma}^\nu)
D_{\mu,\nu}(q)
 ,\label{sigma_gi}\\
{\Gamma}^{\mu}(p_0,q)
&=\gamma\sum_{\boldsymbol{p}}
G(p)
(-e\sigma^{\mu})
G(p-q),\label{gamma_def}
\end{align}
where
$\Gamma_{\mu}(p_0,q)$ is the vertex function shown in Fig.~\ref{fig:Self}(b).
The vertex function can be derived from the Ward-Takahashi's 
identity\cite{Ward1950,Takahashi1957} with the self-energy of impurity scatterings(Eq.~(\ref{life-time})),
\begin{align}
(p'-p)^{\mu}(\sigma_{\mu}+\Gamma_{\mu}(p,p'))=p'-\Sigma^{(i)}(p')-\left(p-\Sigma^{(i)}(p)\right).
\end{align}
The vertex function results in
\begin{align}
\Gamma^{\mu}(p,p')
&=-ie\gamma\delta^{0\mu} , \label{gamma2}
\end{align}
where $\delta^{\mu\nu}$ is the Kronecker's delta.
The correction at another vertex of gauge field (Fig.~\ref{fig:Self} (a)) give the same contribution.
Thus the self-energy in Eq.~(\ref{sigma_gi}) becomes
\begin{align}
\Sigma(p)&=2i\gamma e^2
\sum_{q}
G(p-q)
\frac{1}
{q_0^2-\boldsymbol{q}^2+i\varepsilon}\label{eq.S2}.
\end{align}
The summation of ${q}$ can be done in the same method for deriving Eq.~(\ref{eq.S1}).
By introducing a parameter $x_0=v^2\boldsymbol{p}^2/{p_0}^2$, 
we obtain the expression of the self-energy in the limit of $x_0\rightarrow0$ as
\begin{align}
\lim_{\boldsymbol{p}\rightarrow 0}\Sigma(p_0,\boldsymbol{p})
=&
2\pi e^2\gamma^2\left(
(\mathrm{sgn}(p_0)\sigma^0\label{gap}+\hat{\boldsymbol{p}}\cdot\boldsymbol{\sigma}
\right)
,
\end{align}
where $\hat{\boldsymbol{p}}$ is the unit vector in the direction of $\boldsymbol{p}$.
This is the central result of this paper.
The result does not depend on either 
the magnitude of the energy or that of the momentum. 
From the pole of the scalar denominator of $[G_0(p)^{-1} -\Sigma(p)]$, we obtain 
the energy dispersion of perturbed fermions at $|\boldsymbol{p}|\to 0$ as
\begin{align}
p_0 =&\pm (v|\boldsymbol{p}|+2\pi e^2\gamma^2)+2\pi e^2\gamma^2
\mathrm{sgn}(p_0)\nonumber\\
=&\pm (v|\boldsymbol{p}|+4\pi e^2\gamma^2).\label{spectra}
\end{align}
The result shows the gapped energy spectra around the Dirac node.
We note that both the impurity scatterings and the coupling to the gauge field preserve the TRS.

\section{discussion}

According to the features of the spectra in Eq.~(\ref{spectra}), the magnitude of the gap 
should be independent of temperature. Our results are consistent with
the gapped spectra found in the experiment by T.~Sato {\it{et al.}}~\cite{Sato2011}.
Another experiment\cite{Xu2011} has reported the gapless spectra 
in the similar situation. The energy resolution of experiment in Ref.~\onlinecite{Sato2011} 
seems to be better than that in Ref.~\onlinecite{Xu2011}.

 
We discuss effects of electron-electron interactions via the gauge field on the excitation 
spectra around the Dirac node. 
Because of the dynamics of the gauge field, the interaction potential depends on frequency.
As a result, the interplay between the impurity scatterings and the interaction via the gauge field
opens the gap at the Dirac node.
On the other hand, in condensed matter physics, the electron-electron interactions are usually 
taken into account 
within the static approximation. Namely, the Coulomb interaction acts two fermions at an equal time. 
Therefore such static interaction potential does not depend on the frequency. 
The two different types of interactions in the presence of impurity scatterings may affect 
the excitation spectra in two different ways.
In one-dimension, sufficiently strong static interactions cause the gap~\cite{Xu2006}.
In two-dimension, however, the excitation spectra have been believed to remain gapless even in 
the strong static interactions.
It is necessary to consider the electron-electron interactions via the dynamical boson field
to have the gap at the Dirac node in two-dimension. 
This argument is also important when we consider topological protection of the gapless 
surface state by the bulk-boundary correspondence, which we discuss next.

Finally, it is necessary to discuss a relationship between our results and the bulk-boundary
correspondence of three-dimensional TI in the presence of the TRS.
The $Z_2$ invariant is well defined in (1+3) dimensional fermion space of TI\cite{Fu2007}.
The bulk-boundary correspondence works well to understand the existence of the two-dimensional 
gapless state on its surface. 
In this paper, however, we consider a situation in which the fermion in TI naturally couples 
with the electromagnetic field.
In such situation, the whole physical space consists of the fermion space and the gauge boson one.
Namely, the (1+3) dimensional fermion space is no longer closed independent physical space.
At present, it is unclear if it is possible to apply the topological classification defined solely in the
fermion space to the integrated physical space of the fermion and the gauge boson.
Even in the integrated space,
it may be still possible to consider the topological characterization in the fermion space only.
In such case, the theory would be possible to define some topological invariants
in a similar way as the weak topological invariants~\cite{Roy2009-2}. 
%
Therefore the predictions by the topological theory~\cite{Fu2007,Fu2007-2,Moore2007,Roy2009,Qi2008} in (1+3) 
dimensional fermion space are not always valid. 
This makes the background of the gapped energy spectra appearing at the topologically protected surface state. 
The coupling of fermion with another degree of freedom sometimes changes a topologically non-trivial phase 
to a topologically trivial one.
Our result demonstrates an example of the story.

\section{conclusion}
In conclusion, we have studied effects of the gauge field and the impurity scatterings on 
the energy spectra of massless Dirac fermion at the surface of a three-dimensional
topological insulator. By using the perturbation expansion, we have calculated the self-energy 
around the Dirac node. The interplay between the impurity scatterings and the interaction with 
gauge field leads to the gapped energy dispersion around the Dirac node.
Our results suggest that the translational invariance is also necessary for 
the surface state to be gapless in three-dimensional topological insulators.

\begin{acknowledgments}
This work was supported by 
the "Topological Quantum Phenomena" (No. 22103002) Grant-in Aid for 
Scientific Research on Innovative Areas from the Ministry of Education, 
Culture, Sports, Science and Technology (MEXT) of Japan. 
\end{acknowledgments}

\appendix
\section{Self-energy due to impurity scatterings}\label{AP1}
We derive the self-energy of impurity scattering alone.
Within the Born approximation, the self-energy is given by,
\begin{align}
\Sigma^{(i)}=u^2n_{i}\int d^2\boldsymbol{p}\frac{p_0\sigma^0+v\boldsymbol{p}\cdot\boldsymbol{\sigma}}{{p_0}^2-{v}^2\boldsymbol{p}^2+i\varepsilon}.
\end{align}
The term proportional to momentum $\boldsymbol{p}$ vanishes after integrating of $\boldsymbol{p}$.
The self-energy is calculated as
\begin{align}
\Sigma^{(i)}=&u^2n_{i}\int \frac{d^2\boldsymbol{p}}{(\sqrt{2\pi})^2}\frac{p_0\sigma^0}{{p_0}^2-{v}^2\boldsymbol{p}^2+i\varepsilon}\nonumber\\
=&u^2n_ip_0\sigma^0\nonumber\\
&\times\int  \frac{d^2\boldsymbol{p}}{(\sqrt{2\pi})^2}\int_0^{\infty}\frac{d\alpha}{(-i)}\exp[i\alpha({p_0}^2-{v}^2\boldsymbol{p}^2+i\varepsilon)]\nonumber\\
=&-i\gamma p_0\sigma^0.\\
\gamma=& \pi u^2 n_i /v^2
\end{align}
The result is pure imaginary and gives the inverse of the life time
in agreement with a previous result\cite{Suzuura2002}.
We introduce a dimensionless constant $\gamma$ which contains material parameters.

\begin{widetext} 
\section{Self-energy due to coupling with gauge field}\label{AP2}
Let us calculate the self-energy due to coupling with the gauge field represented by 
Fig.~\ref{fig:SE}(a).
The propagator for a Dirac fermion in the presence of impurity scattering is represented by
\begin{align}
G(p)=\frac{(1+i\gamma)p_0{\sigma}^0+v\boldsymbol{p}\cdot{\boldsymbol{\sigma}}}{(1+i\gamma)^2{p_0}^2-{v}^2{\boldsymbol{p}}^2}.\label{gimp}
\end{align}
By using this propagator, the self-energy in Fig.~\ref{fig:SE}(a) is given by
\begin{align}
\Sigma^{(g)}(p)&=
\sum_{q}\sum_{\mu\nu} 
(-e\sigma^{\mu})
G(p-q)
(-e\sigma_{\nu})D_{\mu\nu}(q)\\  
&=\sum_{q_0,\boldsymbol{q}}\sum_{\mu\nu}
(-e\sigma_{\mu})
\frac{(1+i\gamma)(p_0-q_0)\sigma^0+v(\boldsymbol{p}-\boldsymbol{q})\cdot\boldsymbol{\sigma}}
{{(1+i\gamma)^2(p_0-q_0)}^2-{v}^2(\boldsymbol{p}-\boldsymbol{q})^2}
(-e\sigma^{\nu})
\frac{-g_{\mu\nu}}
{q_0^2-\boldsymbol{q}^2+i\varepsilon}.
\end{align}
The self-energy is expressed 
in the parametric integral representation~\cite{Bergere1974} as,
\begin{align}
\Sigma^{(g)}(p)
=&g_{\mu\nu}\lim_{x\rightarrow0}\int_{-\infty}^{\infty} \frac{dq_0d^2{\boldsymbol{q}}}{(\sqrt{2\pi})^3}\int_0^{\infty}\int_0^{\infty} d{\alpha_1}d{\alpha_2}
(-e\sigma^{\mu})
\frac{1}{i}
\left(\sigma_0\partial_{x_0}-\boldsymbol{\sigma}\cdot\boldsymbol{{\partial}_{x}}\right)
(-e\sigma^{\nu})
\exp\left[
iS
\right],\\
S=&
\alpha_1({(1+i\gamma)^2(p_0-q_0)}^2-{v_F}^2(\boldsymbol{p}-\boldsymbol{q})^2+i\varepsilon)
+\alpha_2({q_0}^2-\boldsymbol{q}^2+i\varepsilon)
+x_0(1+i\gamma)(p_0-q_0)-v\boldsymbol{x}\cdot(\boldsymbol{p}-\boldsymbol{q}),
\end{align} 
where $x_\mu$ and $\alpha_j$ are dummy variables.
The integral with respect to the momenta can be done by the Fresnel integral. The results become
\begin{align*}
\Sigma^{(g)}(p)
=&-e^2\lim_{x\rightarrow0}\int_0^{\infty}\int_0^{\infty}\frac{ d{\alpha_1}d{\alpha_2}}{(\alpha_1v^2+\alpha_2)\sqrt{i(\alpha_1(1+i\gamma)^2+\alpha_2)}}
\frac{1}{i}
\left(\sigma^0\partial_{x_0}+\boldsymbol{\sigma}\cdot\boldsymbol{{\partial}_{x}}\right)
\nonumber\\
&\times
\exp\left[
i\left\{
i(\alpha_1+\alpha_2)\varepsilon
+\frac{\alpha_1\alpha_2(1+i\gamma){p_0}^2}{\alpha_1(1+i\gamma)+\alpha_2}
-\frac{\alpha_1\alpha_2{\boldsymbol{p}}^2}{\alpha_1v^2+\alpha_2}
-\frac{\alpha_2(1+i\gamma)p_0x_0}{\alpha_1(1+i\gamma)^2+\alpha_2}
+\frac{v\boldsymbol{x}\cdot\boldsymbol{p}}{\alpha_1v^2+\alpha_2}
\right\}
\right]
.
\end{align*} 
Next we introduce a integration variable $\rho=\alpha_1+\alpha_2$ and change the 
integration variables as $\alpha_i\rightarrow\rho\alpha_i$. In this way, we obtain
\begin{align}
\Sigma^{(g)}(p)
=&
-e^2
\int_0^{1}\int_0^{1}\frac{d{\alpha_1}d{\alpha_2}}{\alpha_1v^2+\alpha_2}
\frac{\delta(1-\alpha_1-\alpha_2)}{\sqrt{i(\alpha_1(1+i\gamma)^2+\alpha_2)}}
\left(\frac{\alpha_2(1+i\gamma)}{\alpha_1(1+i\gamma)^2+\alpha_2} p_0\sigma^0-\frac{\alpha_2}{\alpha_1v^2+\alpha_2}v\boldsymbol{p}\cdot\boldsymbol{\sigma}\right)
\nonumber\\
&\times
\int_0^{\infty}\frac{d\rho}{\sqrt{i\rho}}
\exp\left[
-\rho\left(
\varepsilon
+\frac{\alpha_1\alpha_2(1+i\gamma)^2}{i(\alpha_1(1+i\gamma)^2+\alpha_2)}{p_0}^2-\frac{\alpha_1\alpha_2}{i(\alpha_1v^2+\alpha_2)}v^2\boldsymbol{p}^2
\right)
\right]
.
\end{align} 
By applying the condition $v^2\ll 1$ and introducing a variable $x_0=v^2\boldsymbol{p}^2/{p_0}^2$, 
the self-energy becomes
\begin{align}
\Sigma^{(g)}(p)
\simeq&-
e^2
\int_0^{1}\frac{d{\alpha}}{\sqrt{\alpha((1+i\gamma)^2{p_0}^2(1-\alpha)-(1+(2i\gamma+\gamma^2)\alpha)v^2\boldsymbol{p}^2}}
\left(\frac{1+i\gamma}{1+2i\gamma\alpha}p_0\sigma^0-\frac{v\boldsymbol{p}\cdot\boldsymbol{\sigma}}{1-\alpha}\right)\\
\simeq&-
\frac{e^2}{\sqrt{{p_0}^2}}(1-i\gamma)
\int_0^{1}\frac{d{\alpha}}{\sqrt{\alpha((1-x_0)-\alpha)}}
\left(
1+\frac{\gamma(2i+\gamma)}{2(1+i\gamma)^2}\frac{1-\alpha}{1-\alpha-x_0}x_0
\right)
\left(\frac{1+i\gamma}{1+2i\gamma\alpha}p_0\sigma^0-\frac{v\boldsymbol{p}\cdot\boldsymbol{\sigma}}{1-\alpha}\right).
\end{align}
We neglect terms proportional to $x_0$ because we focus on the limit of $x_0\simeq0$ in the followings.
The results are
\begin{align}
\Sigma^{(g)}(p)
\simeq&-\frac{e^2}{\sqrt{{p_0}^2}}
\left(
\int_{1-x_0}^{1}\frac{d{\alpha}}{i\sqrt{\alpha(\alpha-(1-x_0))}}
+
\int^{1-x_0}_{0}\frac{d{\alpha}}{\sqrt{\alpha((1-x_0)-\alpha)}}
\right)
\left((1+i\gamma-2i\gamma\alpha)p_0\sigma^0-\frac{v\boldsymbol{p}\cdot\boldsymbol{\sigma}}{1-\alpha}\right)\\
=&-
\frac{e^2}{\sqrt{{p_0}^2}}(1-i\gamma)
\left(
-i(1+i\gamma)p_0\sigma^0\left[\log(2\alpha-(1-x_0)+2\sqrt{\alpha(\alpha-(1-x_0))}) \right]_{1-x_0}^{1}\right.\nonumber\\
&-
(1+i\gamma)p_0\sigma^0\left[\arcsin \frac{-2\alpha+(1-x_0)}{1-x_0} \right]^{1-x_0}_{0}\nonumber\\
&
-2\gamma p_0\sigma^0
\left[\sqrt{\alpha(\alpha-(1-x_0))}-(1-x_0)\log\left|\frac{\sqrt{\alpha-(1-x_0)}-\sqrt{\alpha}}{\sqrt{\alpha-(1-x_0)}+\sqrt{\alpha}}\right| \right]_{1-x_0}^{1}
\nonumber\\
&
-2i\gamma p_0\sigma^0
\left[\sqrt{\alpha((1-x_0)-\alpha)}-(1-x_0)\arcsin \sqrt{\frac{(1-x_0)-\alpha}{1-x_0}} \right]^{1-x_0}_{0}
\nonumber\\
&
+i\frac{v\boldsymbol{p}\cdot\boldsymbol{\sigma}}{\sqrt{x_0}}
\left[\log\frac{(2\alpha-(1-x_0))(\alpha-1)+2x_0-2\sqrt{x_0\alpha(\alpha-(1-x_0))}}{\alpha-1}\right]_{1-x_0}^{1}\nonumber\\
&\left.
-\frac{v\boldsymbol{p}\cdot\boldsymbol{\sigma}}{\sqrt{x_0}}
\left[\arcsin \frac{(-1-x_0)(1-\alpha)+2x_0}{(1-x_0)(1-\alpha)} \right]^{1-x_0}_{0}
\right).
\end{align}
We reach the self-energy 
 as,
\begin{align}
\Sigma^{(g)}(p)
\simeq&-
\frac{e^2}{\sqrt{{p_0}^2}}(1-i\gamma)
\left(
-i(1+i\gamma)p_0\sigma^0\log\frac{1+x_0+2\sqrt{x_0}}{1-x_0}
+\pi p_0\sigma^0
-2\gamma p_0\sigma^0\sqrt{x_0}
+2\gamma p_0\sigma^0(1-x_0)\log\frac{1-\sqrt{x_0}}{1+\sqrt{x_0}}\right.\nonumber\\
&\left.
+i\pi\gamma p_0\sigma_0x_0
-i\frac{v\boldsymbol{p}\cdot\boldsymbol{\sigma}}{\sqrt{x_0}}
\log(1+x_0)
-\pi\frac{v\boldsymbol{p}\cdot\boldsymbol{\sigma}}{\sqrt{x_0}}
\right)
.
\end{align}
Using $x_0=v^2\boldsymbol{p}^2/{p_0}^2\ll 1$, it becomes
\begin{align}
\Sigma^{(g)}(p)
\simeq&-
e^2(1-i\gamma)
\left(
\pi \frac{p_0\sigma^0}{\sqrt{{p_0}^2}}
-\pi\frac{\boldsymbol{p}\cdot\boldsymbol{\sigma}}{\sqrt{\boldsymbol{p}^2}}
-2\gamma\sqrt{x_0} \frac{p_0\sigma^0}{\sqrt{{p_0}^2}}
+2\gamma (1-x_0)\log\left(\frac{1-\sqrt{x_0}}{1+\sqrt{x_0}}\right)
\frac{p_0\sigma^0}{\sqrt{{p_0}^2}}
\right.\nonumber\\
&\left.
-i(1+i\gamma)\frac{p_0\sigma^0}{\sqrt{{p_0}^2}}\log\frac{1+x_0+2\sqrt{x_0}}{1-x_0}
+i\pi\gamma x_0\frac{p_0\sigma^0}{\sqrt{{p_0}^2}}
-i\log(1+x_0)\frac{\boldsymbol{p}\cdot\boldsymbol{\sigma}}{\sqrt{\boldsymbol{p}^2}}
\right)
.
\end{align}
Finally by putting $\gamma\to 0$, it is possible
to obtain the self-energy due to the gauge field alone.
\end{widetext}
\bibliography{TSE}

\end{document}